\documentclass[12pt,a4paper]{article}
\usepackage{amsmath}
\usepackage{amssymb} 
\begin{document}
\begin{titlepage} 
  \begin{flushright} IFUP--TH/2021\\
  \end{flushright} ~\vskip .8truecm 
\begin{center} 
  \Large\bf The continuation method and the real analyticity of the
  accessory parameters: the general elliptic case
\end{center}
\vskip 1.2truecm
\begin{center}
{Pietro Menotti} \\ 
{\small\it Dipartimento di Fisica, Universit{\`a} di Pisa}\\ 
{\small\it 
Largo B. Pontecorvo 3, I-56127, Pisa, Italy}\\
{\small\it e-mail: pietro.menotti@unipi.it}\\ 
\end{center} 
\vskip 0.8truecm
\centerline{March 2021}
                
\vskip 1.2truecm
                                                              
\begin{abstract}
We apply the Le Roy-Poincar\'e continuation method to prove the real
analytic dependence of the accessory parameters on the position of the
sources in Liouville theory in presence of any number of elliptic
sources. The treatment is easily extended to the case of the torus
with any number of elliptic singularities. A discussion is given of
the extension of the method to parabolic singularities and higher
genus surfaces.
\end{abstract}

\end{titlepage}

\section{Introduction}
The accessory parameters appeared first in the Riemann-Hilbert problem
asking for an ordinary differential equation whose solutions transform
according to a given monodromy group \cite{bolibrukh}. They reappear
in Liouville theory in the quest for an auxiliary differential
equation in which all elements of the monodromy group belong to
$SU(1,1)$. Such a request is the necessary and sufficient condition
for having a single valued conformal Liouville field. Their
determination also play a crucial role in $2+1$ dimensional gravity
\cite{CMS2} in presence of matter.  This is also connected to the
Polyakov relation which relates such accessory parameters to the
variation of the on-shell action of Liouville theory under the change
in the position of the sources \cite{ZT1,ZT2,CMS1,CMS2,TZ}. They
appear again in the classical limit of the conformal blocks of the
quantum conformal theory
\cite{ZZ,HJP,FP,menottiAccessory,piatek,LLNZ,
  menottiConformalBlocks,menottiTorusBlocks}.

In several developments it is important to establish the nature of the
dependence of such accessory parameters on the source positions and on
the moduli of the theory. To this end we have the result of Kra
\cite{kra} which in the case of the sphere topology in presence only
of parabolic and finite order elliptic singularities proved that such
a dependence is real analytic (not analytic). The technique used to
reach such a result was that of the fuchsian mapping, a method which
cannot be applied to the case of general elliptic singularities.

On the other hand in the usual applications, general elliptic, not
finite order elliptic singularities appear. Finite order singularities
are those for which the source strength is given by $\eta_k
=(1-1/n)/2, n\in Z_+$ (see section 2).

In the case when only one independent accessory parameter is present,
like the sphere topology with four sources or the torus with one
source, it was proven that such accessory parameters are real analytic
functions of the source position or moduli, almost everywhere
(i.e. everywhere except for a zero measure set) in the source position
or moduli space
\cite{menottiAccessory,
  menottiPreprint,menottiHigherGenus,menottiTorusBlocks}. The
qualification almost everywhere implies e.g. that we could not exclude
the presence of a number of cusps in the dependence of the accessory
parameters on the source positions, a phenomenon which may be expected
in the solution of a system of implicit equations.

This result was obtained by applying complex variety techniques to the
conditions which impose the $SU(1,1)$ nature of all the monodromies.
In \cite{menottiPreprint} an extension of such a technique was attempted
to the case of two independent accessory parameters, like the sphere
with five sources and the torus with two sources but results where
obtained only under an irreducibility assumption.

The usual approach to the solution of the Liouville equation is the
variational approach. Such an approach was suggested by Poincar\'e in
\cite{poincare} but not pursued by him due to some difficulties in
proving the existence of the minimum of a certain functional. The
variational approach was developed with success by Lichtenstein
\cite{lichtenstein} and in a different context by Troyanov
\cite{troyanov} by writing the conformal field as the sum of a proper
background and a remainder. With such a splitting the problem is
reduced to the search of the minimum of a given functional. One proves
that the such a minimum exists and solves the original problem
\cite{lichtenstein,troyanov,menottiExistence}. Poincar\'e in
\cite{poincare} pursued and solved the same problem by means of a
completely different procedure which became known as the Le
Roy-Poincar\'e continuation method \cite{leroy,mahwin}.
 
The idea is to write the solution of the Liouville equation as a power
series expansion in certain properly chosen parameters. Such a series
turns out to be uniformly convergent over all the complex plane or
Riemann surface.

This cannot be achieved in a single step. Once one has solved the
equation with one of such parameter in a certain region one uses the
obtained solution as the starting point of an other series in another
parameter and thus at the end one has the solution as a series of
series, each uniformly convergent.

The procedure is more lengthy than the variational approach but has
the advantage that one can follow the dependence of each series on the
input, the input being the Lichtenstein background field.

Such a field, to be called $\beta$ is a real positive function smooth
everywhere except at the source positions, the singularity being
characterized by the nature and the strength of the sources; apart
from these requirements the choice of $\beta$ is free. Thus except at
the singularities $\beta$ is a smooth, say $C^\infty$ function. The
uniqueness theorem \cite{lichtenstein,menottiExistence} tells us that
the final result does not depend on the specific choice of $\beta$.

Simple smoothness would not be a good starting point for proving the
real analytic dependence of the result; on the other hand as we
shall see, it is
possible to provide a background field $\beta$ satisfying all the
Lichtenstein requirements and real analytic in the moduli except
obviously at the sources.

Starting from such a $\beta$ one sees that the zero order
approximation in the Poincar\'e procedure gives rise to a conformal
field which is real analytic in the position of the sources $z_k$ and
in the argument $z$ except at the source positions $z_k$. The problem
is to show that such real analyticity properties are inherited in all
power expansion procedures and finally by the conformal factor itself.

This is what is proved in this paper in presence on any number of
elliptic singularities. The final outcome is that the
conformal factor depends in real analytic way both on the argument $z$
of the field and on the source positions. Once this result is
established is not difficult to express the accessory parameters in
terms of the conformal field and prove the real analytic dependence of
the accessory parameters themselves on the source positions.

The paper is structured as follows. In section \ref{lichtenstein} we
describe the Lichtenstein decomposition and provide a background field
$\beta$ which is real analytic everywhere except at the sources.

In section \ref{poincareprocedure} we give the Poincar\'e procedure
for the solution of the Liouville equation and in the following
section \ref{linearsection} we give the method of solution for a class
of linear inhomogeneous equation which appear in section
\ref{poincareprocedure}. In section \ref{inheritance} we prove how the
real analytic properties of the background field $\beta$ are inherited
in all the iteration process and finally by the solution i.e. the
Liouville field. In section \ref{realanalyticitysection}, using the
obtained result we prove the real analytic dependence of the accessory
parameters on the source positions for the sphere topology for any
number of general elliptic singularities.  In section
\ref{torusanalyticity} we give the extension of the result to the
torus with any number of sources.

Finally in section \ref{conclusions} we discuss the perspectives
for the extension of the method to the parabolic singularities and
higher genus. For making the paper more readable we have relegated in
four appendices the proof of some technical results which are employed
in the text.

\section{The Lichtenstein decomposition}\label{lichtenstein}

The Liouville equation is
\begin{equation}
\Delta\phi=e^\phi
\end{equation}
with the boundary conditions at the elliptic singularities
\begin{equation}
  \phi+2\eta_k \log|z-z_k|^2={\rm bounded},~~~~~~~\eta_k<\frac{1}{2}
\end{equation}
and at infinity
\begin{equation}
  \phi+2\log|z|^2={\rm bounded}~~.
\end{equation}
The procedure starts by constructing a positive function $\beta$
everywhere smooth except at the sources where it obeys the
inequalities
\begin{equation}\label{inequality1}
0<\lambda_m<\beta |z-z_k|^{4\eta_k} <\lambda_M
\end{equation}
and for $|z|>\Omega$, $\Omega$ being the radius of a disk which
include all singularities
\begin{equation}\label{inequality2}
0<\lambda_m<\beta |z|^4<\lambda_M ~. 
\end{equation}
Note that $\int \beta(z) d^2z<\infty$. In addition $\beta$ will be normalized
as to have
\begin{equation}\label{sumrule}
-\sum_k2\eta_k+\frac{1}{4\pi}\int\beta(z)d^2z=-2
\end{equation}
for the sphere topology.

Apart from these requirements  $\beta$ is free and due
to the  uniqueness theorem the final result for the field $\phi$ does
not depend on the specific choice of $\beta$. On the other hand, as
discussed in the introduction, it will be useful to start from a
$\beta$ which is real analytic both in $z$ and in the source positions
$z_k$, except at the sources. One choice is
\begin{equation}\label{ourbeta}
\beta = c \prod_k\frac{[(z-z_k)(\bar z-\bar
    z_k)]^{-2\eta_k}}{[1+z\bar z]^{-2\sigma+2}},~~~~~~~~\sigma=\sum_k \eta_k
\end{equation}
where the positive constant $c$ has to be chosen as to comply with
the sum rule (\ref{sumrule}).
Picard inequalities
require the presence of at least three singularities, the case of
three singularities being soluble in terms of hypergeometric functions.
As is well known, by performing a projective
transformation we can set $z_1=0, z_2=1, z_3=i$.
We shall be interested in the dependence
of the accessory parameters on a given $z_k$ keeping the other fixed;
we shall call such a source position $z_4$. 
Obviously (\ref{ourbeta}) is not the only choice but it is
particularly simple. Varying the position $z_4$ around a given initial
position we shall need to vary the $c$ in
order to keep (\ref{sumrule}) satisfied. It is easily seen that
such $c$ depends on $z_4$ in  real analytic way (see Appendix A).

Given $\beta$ one constructs the function
\cite{lichtenstein,menottiExistence}
\begin{equation}\label{liouville}
\nu = \phi_1+\frac{1}{4\pi}\int\log|z-z'|^2\beta(z')d^2z'\equiv\phi_1+I
\end{equation}
with
\begin{equation}
\phi_1 =\sum_k(-2\eta_k) \log|z-z_k|^2
\end{equation}
and we define $u$ by
\begin{equation}
\phi = \nu+u~.
\end{equation}
With such a definition the Liouville equation becomes
\begin{equation}\label{liouville2}
\Delta u = e^\nu e^u-\beta \equiv \theta e^u-\beta~.
\end{equation}

The real analyticity of $\beta$ and $\theta$ need a little discussion.
We recall that a real analytic function can be defined as the value
assumed by an analytic function of two variables $f(z,z^c)$ when $z^c$
assumes the value $\bar z$. Equivalently it can be defined as a
function of two real variables $x$ and $y$ which locally can be
expanded in a convergent power series
\begin{equation}
  f(x+\delta x,y+\delta y) - f(x,y)= \sum_{m,n} a_{m,n} \delta x^m
  \delta y^n~.
\end{equation}
In eq.(\ref{ourbeta}) we can write
\begin{equation}
  [(z-z_k)(\bar z-\bar z_k)]^{-2\eta_k}=
  [(x-x_k)^2+(y-y_k)^2]^{-2\eta_k}
\end{equation}
which around a point $x,y$ with $x\neq x_k$ and/or $y\neq y_k$
can be expanded in a power series, obviously with bounded convergence
radius. The function $\nu$ and consequently the function $\theta$
contain $\beta$ in the form
\begin{equation}
  e^\nu = e ^{\phi_1+\frac{1}{4\pi}\int\log|z-z'|^2\beta(z')d^2z'}=
  \prod_k [(z-z_k)(\bar z-\bar z_k)]^{-2\eta_k} ~e^I.
\end{equation}
As we shall keep all $z_k$ fixed except $z_4$ we shall write
\begin{equation}
I(z,z_4)=\frac{1}{4\pi}\int\log|z-z'|^2\beta(z',z_4)d^2z'
\end{equation}
The analytic properties of $I$ both in $z$ and $z_4$ are worked out
in Appendix B.

\section{The Poincar\'e procedure}\label{poincareprocedure}

After performing the decomposition of the field $\phi$ as
$\phi=u+\nu$ the Liouville equation becomes
\begin{equation}\label{originaleq}
\Delta u= \theta e^u-\beta~~~~~{\rm with}~~\theta=e^\nu\equiv r\beta 
\end{equation}
and as a consequence of the inequalities (\ref{inequality1},
\ref{inequality2}) we have
\begin{equation}\label{r1rr2}
0<r_1<r<r_2
\end{equation}
for certain $r_1,r_2$.

Let $\alpha$ be the minimum
\begin{equation}
\alpha=\min\bigg(\frac{\beta}{\theta}\bigg)=\frac{1}{\max r}
\end{equation}
which due to (\ref{r1rr2}) is a positive number.
Then we can rewrite the equation as
\begin{equation}\label{rewritten}
\Delta u = \theta e^u - \alpha\theta -\beta(1-\alpha r)~.
\end{equation}
As a consequence of the choice for $\alpha$ we have $\psi\equiv
\beta(1-\alpha r)\geq 0$.

Convert the previous equation to 
\begin{equation}\label{lambdaeq}
  \Delta u = \theta e^u-\alpha\theta -\lambda\psi
\end{equation}
and write
\begin{equation}\label{nonlinearseries}
  u=u_0+\lambda u_1+\lambda^2 u_2+\dots~.
\end{equation}  
We have to solve the system 
\begin{eqnarray}\label{nonlinearsystem}
  &&\Delta u_0 = \theta (e^{u_0}-\alpha)\nonumber\\
  &&\Delta u_1 = \theta e^{u_0}u_1-\psi\nonumber\\
  &&\Delta u_2 = \theta e^{u_0}(u_2+w_2)\nonumber\\
  &&\Delta u_3 = \theta e^{u_0}(u_3+w_3)\nonumber\\
  &&\dots
\end{eqnarray}  
where
\begin{equation}
  w_2 = \frac{u_1^2}{2},~~~~w_3 = \frac{u_1^3}{6}+u_1u_2,~~~~
  w_4 = \frac{u_1^4}{24}+\frac{u_1^2 u_2+u_2^2}{2}+u_1u_3,~~~~\dots
\end{equation}  
are all polynomials with positive coefficients. We see that in
the $n$-th equation the $w_n$ is given in terms of $u_k$ with $k<n$
and thus each of the equations (\ref{nonlinearsystem}) is a linear
equation.

Thus the previous is a system of linear inhomogeneous differential
equation for the $u_k$. The first equation is solved by $u_0=\log
\alpha$.  We shall see in the next section that each of the following
equations in (\ref{nonlinearsystem}) can be solved by iterated power
series expansion and that all the $u_k$ are bounded. From the
properties of the Laplacian $\Delta$ and eq.(\ref{nonlinearsystem}) we
have
\begin{eqnarray}\label{inequalities}
  &&|u_1| \leq \max\bigg(\frac{\psi}{e^{u_0}\theta}\bigg)\nonumber\\
   &&|u_2| \leq \max~|w_2|\nonumber\\
  &&\dots\nonumber\\
  &&|u_k| \leq \max~|w_k|\nonumber\\
  && \dots
\end{eqnarray}
If at $z=z_{\rm max}$, where $z_{\rm max}$ is the point where $|u_k|$
reaches its maximum, $\Delta u_k$ is finite the above inequalities
follow from the well known properties of the Laplacian. At the
singular points it may happen that the Laplacian diverges but the
inequalities (\ref{inequalities}) still hold. In fact if the maximum
of $|u_k|$ is reached at the singular point $z_l$, with $u_k(z_l)>0$
and the r.h.s. in eq.(\ref{nonlinearsystem}) is definite positive in a
neighborhood of $z_l$, then the circular average
\begin{eqnarray}
  \frac{1}{2\pi}\int u_k(z_l+\rho e^{i\varphi})d\varphi \equiv \bar u(\rho)
\end{eqnarray}
has a positive definite source. Thus it is increasing with $\rho$
which contradicts the fact that $u_k(z_l)$ is the maximum. The same
reasoning works also if at $z_l$ we have $u_k(z_l)<0$.

Using the above inequalities one proves (see Appendix D) 
that the series (\ref{nonlinearseries}) converges for
\begin{equation}\label{lambda0bound}
  |\lambda|<\frac{\alpha(\log 4-1)}
  {\max|\frac{\psi}{\theta}|}
\end{equation}
and such convergence is uniform.

It is not difficult to show \cite{poincare} using the results of
Appendix C, that the convergent series satisfies the differential
equation (\ref{lambdaeq}) i.e. that one can exchange in
(\ref{lambdaeq}) the Laplacian with the summation operation.

Thus we are able to solve the equation
\begin{equation}
\Delta u = \theta e^u - \alpha\theta -\lambda_0\psi
\end{equation}
for
\begin{equation}
0<\lambda_0<\frac{\alpha(\log4-1)}{\max~|\frac{\psi}{\theta}|}~.
\end{equation}
If $\lambda_0$ can be taken equal to $1$ the problem is solved.
Otherwise one can extend the region of solubility of our equation by
solving the equation
\begin{equation}
  \Delta u = \theta e^u - \theta \alpha-
    \lambda_0\psi
  -\lambda\psi\equiv
  \theta e^u - \varphi-\lambda\psi~.
\end{equation}
Expanding as before in $\lambda$ one obtains
\begin{eqnarray}
  &&\Delta u_0 = \theta e^{u_0}-\varphi\nonumber\\
  &&\Delta u_1 = \theta e^{u_0}u_1-\psi\nonumber\\
  &&\Delta u_2 = \theta e^{u_0}(u_2+w_2)\nonumber\\
  &&\Delta u_3 = \theta e^{u_0}(u_3+w_3)\nonumber\\
  &&\dots
\end{eqnarray}  
From the first equation using $\varphi>0$ we have
\begin{equation}
\min~e^{u_0} >\min(\frac{\varphi}{\theta})  
\end{equation}
and thus from the second
\begin{equation}
  \max~ |u_1| \leq\max |\frac{\psi}{\theta e^{u_0}}|\leq
  \max |\frac{\psi}{\theta}|\frac{1}{\min(\frac{\varphi}{\theta})}=
  \max|\frac{\psi}{\theta}|\frac{1}{\min(\alpha+\lambda_0\frac{\psi}{\theta})}<
  \frac{\max~|\frac{\psi}{\theta}|}{\alpha}~.
\end{equation}
Then following the procedure of the previous step we have convergence
for
\begin{equation}
|\lambda|<\frac{\alpha(\log 4-1)}{\max~|\frac{\psi}{\theta}|}~.
\end{equation}  
This is the same bound as (\ref{lambda0bound}) and thus repeating such
extension procedure, in a finite number of steps we reach the solution
of the original equation (\ref{originaleq}).  We shall call these
steps extension steps.

\section{The equation $\Delta u = \eta u -\varphi$}
\label{linearsection}
In the previous section we met the problem of solving linear
equations in $u$ of the type
\begin{equation}\label{linearequation}
\Delta u = \theta e^U u -\varphi
\end{equation}
where $U$ is provided by the solution of a previous equation.  Here we
give the procedure for obtaining the solution of the more general
equation
\begin{equation}\label{generallinearequation}
\Delta u = \eta u -\varphi
\end{equation}
where $\eta$ is positive and has
the same singularities as $\theta$ in the sense that $
0<c_1<\frac{\eta}{\theta}<c_2$ \cite{poincare}. We start noticing
that due to the positivity of $\eta$, if $u$ and $v$ are two solutions of
(\ref{generallinearequation}) then we have
\begin{equation}
  \int(u-v)\Delta (u-v)d^2z=-\int\nabla(u-v)\cdot\nabla(u-v)d^2z=
  \int\eta(u-v)^2d^2z=0
\end{equation}
i.e. $u=v$.  To construct the solution one considers the equation
\begin{equation}\label{lambdadiffeq}
\Delta u = \lambda\eta u -\varphi_0 -\lambda\psi
\end{equation}
with $\int\varphi_0 d^2z=0$ and writes the $u$ as
\begin{equation}\label{ulinearexp}
u = (u_0+c_0)+\lambda(u_1+c_1)+\lambda^2(u_2+c_2)+\dots
\end{equation}
and then we have
\begin{eqnarray}\label{linearsystem}
  &&\Delta u_0 = -\varphi_0 \nonumber\\
  &&\Delta u_1 = \eta (u_0+c_0)-\psi\nonumber\\
  &&\Delta u_2 = \eta (u_1+c_1)\nonumber\\
  &&\Delta u_3 = \eta (u_2+c_2)\nonumber\\
  &&\dots
\end{eqnarray}
where the $u_k$ are simply given by
\begin{equation}
u_k=\frac{1}{4\pi}\int\log|z-z'|^2 s_k(z')d^2z'
\end{equation}  
being $s_k$ the sources in eq.(\ref{linearsystem}). Due to the compactness
of the domain, i.e. the Riemann sphere, equations of the type $\Delta u=
s$ are soluble only if $\int s d^2z=0$.  
The solutions of
the $\Delta u = s$ are determined up to a constant, a fact which
have been explicitly taken into account in (\ref{ulinearexp}).

Then the $c_k$ are chosen as to have the integral of the r.h.s. of the
equations in (\ref{linearsystem}) equal to zero.

\begin{eqnarray}\label{cequations}
  &&c_0\int\eta d^2z= \int \psi d^2z - \int\eta u_0 d^2z\nonumber\\
  &&c_1\int\eta d^2z= - \int\eta u_1 d^2z\nonumber\\
  &&c_2\int\eta d^2z= - \int\eta u_2 d^2z\nonumber\\
  && \dots
\end{eqnarray}  
Thus we have $|c_k|<\max|u_k|$ for $k\geq 1$.  On the other hand we
have from the inequality proven in Appendix B
\begin{equation}
\max|u_2|\leq B \max~|u_1+c_1|
\end{equation}
from which
\begin{equation}
\max~|u_2+c_2|\leq 2 \max |u_2| < 2B \max~|u_1+c_1|
\end{equation}
and similarly for any $k$. Thus the series converges uniformly for
$|\lambda|<\frac{1}{2B}$.  Again one can easily prove \cite{poincare}
using the results of Appendix C, that one can exchange the summation
operation with the Laplacian and thus the series satisfies the
differential equation (\ref{lambdadiffeq}).
It is important to notice
that the convergence radius does not depend on $\varphi$.
Then chosen any $\lambda_1$, $0<\lambda_1< \frac{1}{2B}$ we can solve
for any $\varphi$
\begin{equation}\label{generallinear}
\Delta u =\lambda_1 \eta u -\varphi\equiv\lambda_1 \eta u -\varphi_0-\psi
\end{equation}
as the power expansion in $\lambda$ of
\begin{equation}
\Delta u =\lambda \eta u -\varphi_0 - \frac{\lambda}{\lambda_1}\psi
\end{equation}
converges for $\lambda=\lambda_1$.

Thus if $\frac{1}{2B}>1$ the problem is solved.  Otherwise one can
extend the region of convergence in the following way.

Chosen $0<\lambda_1 = \frac{1}{2B}-\varepsilon$
we consider the equation
\begin{equation}
  \Delta u=\lambda_1 \eta u+\lambda \eta u -\varphi~.
\end{equation} 
We are already able to solve
\begin{equation}
\Delta u=\lambda_1\eta u -\varphi
\end{equation}  
and thus we shall expand in $\lambda$
\begin{equation}
u=u_0+\lambda u_1+\lambda^2 u_2+\dots
\end{equation}  
with
\begin{eqnarray}\label{seconditeration}
  &&\Delta u_0=\lambda_1\eta u_0-\varphi\nonumber\\
  &&\Delta u_1=\lambda_1\eta u_1+\eta u_0\nonumber\\
  &&\Delta u_2=\lambda_1\eta u_2+\eta u_1\nonumber\\
  &&\Delta u_3=\lambda_1\eta u_3+\eta u_2\nonumber\\
  &&\dots
\end{eqnarray}  
all of which are of the form (\ref{generallinear}) and thus we are
able to solve.

To establish the convergence radius in $\lambda$ we use the fact
that in the solution of (\ref{seconditeration}) we have
\begin{equation}
|u_{k+1}|<\frac{1}{\lambda_1}\max|u_{k}|,~~~~~~~~k\geq 1
\end{equation}  
and thus we have uniform convergence of the series in $\lambda$ for
$|\lambda|<\lambda_1$.  We repeat now the procedure starting from
the equation
\begin{equation}
  \Delta u=\lambda_1 \eta u+\lambda_2 \eta u +\lambda \eta u
  -\varphi
\end{equation}
with $0<\lambda_2<\lambda_1$ which is solved again by expanding in
$ \lambda$. From the same
argument as before the convergence radius in $\lambda$ is
\begin{equation}
\lambda_1+\lambda_2
\end{equation}
which is even larger than the convergence radius in $\lambda_2$ and
thus in a finite number of extension steps we are able to solve
\begin{equation}
\Delta u=(\lambda_1+\lambda_2+\dots+\lambda_n)\eta u -\varphi
\end{equation}  
with $\lambda_1+\lambda_2+\dots+\lambda_n=1$ which is our original
equation (\ref{generallinearequation}).

\section{The inheritance of real analyticity}\label{inheritance}

Not to overburden the notation we shall write $f(z,z_4)$ for
$f_c(z,z^c,z_4,z_4^c)$ at $z^c=\bar z$ and $z_4^c=\bar z_4$ with
$\frac{\partial}{\partial z}=\frac{1}{2}(\frac{\partial}{\partial x}
-i\frac{\partial}{\partial y})$ and
$\frac{\partial}{\partial \bar z}=\frac{1}{2}(\frac{\partial}{\partial
  x} +i\frac{\partial}{\partial y})$.

We need the detailed structure of the most important function which
appears in the iteration procedure i.e. of $\theta=\beta r$. We are
interested in the problem when $z_4$ varies in a domain $D_4$ around a
$z^0_4$, say $|z_4-z^0_4|<R_4$ which excludes all others singularities.
We choose $R_4$ equal to $1/4$ the minimal distance of $z_4^0$ from
the singularities $z_k$, $k\neq 4$.  We know that $0<r_1<r<r_2$ where
the bounds $r_1$ and $r_2$ can be taken independent of $z_4$ for
$z_4\in D_4$.  The function $\theta(z,z_4)$ is explicitly given by
\begin{equation}
\theta = \prod_k ((z-z_k)(\bar z-\bar z_k))^{-2\eta_k} e^{I(z,z_4)}
\end{equation}
where
\begin{equation}\label{I}
I(z,z_4)=\frac{1}{4\pi}\int\log|z-z'|^2 \beta(z',z_4)d^2z'~.
\end{equation}
In dealing with integrals of the type (\ref{I}) to avoid the
appearance of non integrable functions in performing the derivative
w.r.t. $z_4$ it is instrumental to isolate a disk ${\cal R}_1$ around
$z_4$ of radius $R_1$ that for $z_4\in D_4$ contains only the
singularity $z_4$ and not the others $z_k$.

For the function $u$ of $z$ and $z_4$ it is useful to write for
$|z-z_4|<R_1$, $u(z,z_4)=\hat u(\zeta,z_4)$ with $\zeta=z-z_4$ and
thus also $\hat\theta(\zeta,z_4)=\theta(z,z_4)$ for $|\zeta|<
R_1$. Thus for $|z-z_4|<R_1$ we shall have denoting with ${\cal
  R}_{1c}$ the complement of ${\cal R}_1$
\begin{eqnarray}\label{exphatI}
&& \hat I(\zeta,z_4) =\frac{1}{4\pi}\int_{{\cal R}_1}\log|\zeta-\zeta'|^2
  \hat\beta(\zeta',z_4)
  d^2\zeta'\nonumber\\
  &&+\frac{1}{4\pi}\int_{{\cal R}_{1c}}\log|\zeta+z_4-z'|^2
  \beta(z',z_4) d^2z'~.
\end{eqnarray}
We shall also consider an other disk centered in $z_4$ with
radius $R_2<R_1$ and write for $|z-z_4|>R_2$
\begin{eqnarray}\label{expI}
&& I(z,z_4) =\frac{1}{4\pi}\int_{{\cal R}_2}\log|z-z_4-\zeta'|^2
  \hat\beta(\zeta',z_4)
  d^2\zeta'\nonumber\\
  &&+\frac{1}{4\pi}\int_{{\cal R}_{2c}}\log|z-z'|^2
  \beta(z',z_4) d^2z'~.
\end{eqnarray}
In Appendix B it is proven that $I(z,z_4)$ eq.(\ref{I}), is continuous
and it is real analytic in $z$ for $z\neq z_k$ and that $\hat I(\zeta,z_4)$
eq.(\ref{exphatI}) is real analytic in $z_4$ for $z_4\in D_4$ and
$I(z,z_4)$ eq.(\ref{expI}) real analytic in $z_4\in D_4$.

\bigskip

The typical transformation we where confronted with in the previous
sections was
\begin{equation}\label{transformation}
u(z,z_4)=\frac{1}{4\pi}\int\log|z-z'|^2 \theta(z',z_4) s(z',z_4)d^2z'~.
\end{equation}
For $|z-z_4|<R_1$ we have
\begin{eqnarray}\label{transfin}
 && \hat u(\zeta,z_4) =
  \frac{1}{4\pi}\int_{{\cal R}_1}\log|\zeta-\zeta'|^2
  \hat\theta(\zeta',z_4) \hat s(\zeta',z_4)d^2\zeta'\nonumber\\
&&+  \frac{1}{4\pi}\int_{{\cal R}_{1c}}\log|\zeta+z_4-z'|^2
  \theta(z',z_4) s(z',z_4)d^2z'
\end{eqnarray}
and for $|z-z_4|> R_2$ we have
\begin{eqnarray}\label{transfout}
 && u(z,z_4) =
  \frac{1}{4\pi}\int_{{\cal R}_2}\log|z-\zeta'-z_4|^2
  \hat\theta(\zeta',z_4) \hat s(\zeta',z_4)d^2\zeta'\nonumber\\
&&+  \frac{1}{4\pi}\int_{{\cal R}_{2c}}\log|z-z'|^2
  \theta(z',z_4) s(z',z_4)d^2z'~.
\end{eqnarray}
We recall that we work under the condition
\begin{equation}\label{integral0}
\int\theta(z',z_4) s(z',z_4)d^2z'=0
\end{equation}
which can also be written as
\begin{equation}
  \int_{\cal R}\hat\theta(\zeta',z_4) \hat s(\zeta',z_4)d^2\zeta'+
  \int_{{\cal R}_c}\theta(z',z_4) s(z',z_4)d^2z'=0~.
\end{equation}
A consequence of relation (\ref{integral0}) is that we can work also
with
\begin{eqnarray}
 && \hat u(\zeta,z_4) =
  \frac{1}{4\pi}\int_{{\cal R}_1}\log\big|1-\frac{\zeta'}{\zeta}\big|^2
  \hat\theta(\zeta',z_4) \hat s(\zeta',z_4)d^2\zeta'\nonumber\\
&& + \frac{1}{4\pi}\int_{{\cal R}_{1c}}\log\big|1+\frac{z_4-z'}{\zeta}\big|^2
  \theta(z',z_4) s(z',z_4)d^2z'~,
\end{eqnarray}

\begin{eqnarray}\label{u1out}
 && u(z,z_4) =
  \frac{1}{4\pi}\int_{{\cal R}_2}\log\big|1-\frac{\zeta'+z_4}{z}\big|^2
  \hat\theta(\zeta',z_4) \hat s(\zeta',z_4)d^2\zeta'\nonumber\\
&&+  \frac{1}{4\pi}\int_{{\cal R}_{2c}}\log\big|1-\frac{z'}{z}\big|^2
  \theta(z',z_4) s(z',z_4)d^2z'~.
\end{eqnarray}
This last form is useful in investigating the behavior of
$u(z,z_4)$ at $z=\infty$.

\bigskip

We shall now show that some boundedness and real analyticity
properties of the source $s(z,z_4)$ are inherited by $u(z,z_4)$
through the transformation (\ref{transformation}).  We shall always
work with $z_4\in D_4$ where $D_4$ was described at the beginning of
the present section and does not contain any other singularity $z_k$.
The real analyticity is proven by showing the existence the complex
derivatives w.r.t. $z$ and $\bar z$ or w.r.t $z_4$ and $\bar z_4$.
Due to the symmetry of the problem it is sufficient to prove
analyticity w.r.t. $z$ and $z_4$.

Properties of the source $s$ which are inherited by $u$ in the
transformation (\ref{transformation}) are

\bigskip

P1. $u$ is bounded and continuous in $z,z_4$, $z_4\in D_4$

P2. $\hat u(\zeta,z_4)$ is analytic in $\zeta$ for $|\zeta|<R_1$,
$\zeta\neq 0$.
 
P3. $\hat u(\zeta,z_4)$ is analytic in $z_4$  with
$\frac{\partial \hat u(\zeta,z_4)}{\partial z_4}$ bounded for
$z_4\in D_4$, $|\zeta|<R_1$.

P4. $u(z,z_4)$ is analytic in $z$, for $|z-z_4|>R_2$ , $z=\infty$
included, except at $z=z_k$.

P5. $u(z,z_4)$ is analytic in $z_4$  with
$\frac{\partial u(z,z_4)}{\partial z_4}$ bounded for
$z_4\in D_4$, $|z-z_4|>R_2$.

\bigskip

Thus we shall assume that the properties P1-P5 are satisfied by
$s(z,z_4)$ and prove that they are inherited by $u(z,z_4)$ of
eq.(\ref{transformation})

\bigskip

The inheritance of property P1 is a consequence of the inequality
proven in Appendix C.

The inheritance of properties P2 and P4 is proved by computing the
derivative w.r.t. $z$ using the method employed in Appendix B when
dealing with the derivative of $I(z,z_4)$ and using the analyticity
and boundedness of $s(z,z_4)$.

As for P3 we shall use the expression (\ref{transfin}) for $\hat
u$. $\hat\theta$ has the following structure

\begin{equation}\label{thetahat}
  \hat\theta(\zeta,z_4) = (\zeta\bar\zeta)^{-2\eta_4} \prod_{k\neq 4}
  (|\zeta+z_4-z_k|^2)^{-2\eta_k} e^{\hat I(\zeta,z_4)},~~~~~~~~|\zeta |<R_1~.
\end{equation}  

With respect to the first term in (\ref{transfin}), in taking the
derivative w.r.t. $z_4$ one easily sees that the conditions are
satisfied for taking the derivative under the integral sign, for all
$\zeta$, provided $\hat s$ and
$\frac{\partial \hat s}{\partial z_4}$ be bounded in ${\cal R}_1\times
D_4$, i.e. properties P1 and P3.
 
In fact the derivative of the product in eq.(\ref{thetahat})
w.r.t. $z_4$ is regular and the derivative of the exponential boils
down to the derivative of $\hat I$ which we have shown in Appendix B
to be analytic in $z_4$ for all $\zeta$ in ${\cal R}_1$.

Then we have to differentiate $\hat s(\zeta,z_4)$ w.r.t.  $z_4$. As
the $\frac{\partial\hat s}{\partial z_4}$ uniformly bounded in ${\cal
  R}_1$, property P3, such differentiation under the integral sign is
legal.

In taking the derivative w.r.t. $z_4$ of the second term in
(\ref{transfin}) we must take into account the fact that the
integration region ${\cal R}_{1c}$ moves as $z_4$ varies. Then the
derivative of the second integral appearing in (\ref{transfin}) is
\begin{eqnarray}
&&  \frac{1}{4\pi}\int_{{\cal R}_{1c}} \frac{\partial}{\partial z_4}
  \bigg[\log|\zeta+z_4-z'|^2 
    \theta(z',z_4) s(z',z_4)\bigg] d^2z'\nonumber\\
&-& \frac{1}{8\pi i}\oint_{\partial {\cal R}_{1c}} \log|\zeta+z_4-z'|^2
  \theta(z',z_4) s(z',z_4)  d\bar z'~.
\end{eqnarray}
The logarithms in the above equation are not singular for
$\zeta\in{\cal R}_1$ which makes the differentiation under
the integral sign legal.

We come now to the $u(z,z_4)$ with $|z-z_4|>R_2$ where we use
expression (\ref{transfout}). In taking the derivative w.r.t. $z_4$
the first integral does not present any problem as $z\in {\cal R}_{2c}$
and $\zeta\in {\cal R}_2$ and thus the logarithm is non singular.  The
second integral gives two contributions, being the first provided by
the derivative of $\theta s$ which gives rise to an integrand bounded
by an absolutely integrable function, independent of $z_4$ for $z_4\in
D_4$ and a contour integral due to the motion of ${\cal R}_{2c}$ as $z_4$
varies.

We are left to examine the neighborhood of $z=\infty$ which, with the
behavior (\ref{inequality2}) for the $\beta$ at infinity and the
consequent behavior of the $\theta$, is a regular point.  We have with
$\tilde u(x,z_4)=u(1/x,z_4)$ and $\tilde\theta(y,z_4)=\theta(1/y,z_4)$
and using (\ref{u1out})
\begin{equation}\label{tildeequation}
  \tilde u(x,z_4)=\frac{1}{4\pi}\int\log|x-y|^2\frac{\tilde\theta(y,z_4)}
         {(y\bar y)^2}\tilde s(y,z_4)d^2y-
         \frac{1}{4\pi}\int\log|y|^2\frac{\tilde\theta(y,z_4)}
  {(y\bar y)^2}\tilde s(y,z_4)d^2y~.
\end{equation}
Exploiting the analyticity of $\tilde\theta(y,z_4)/(y\bar y)^2$ for
$|y|<1/\Omega$ we have that (\ref{tildeequation}) has complex
derivative w.r.t. $x$ for $|x|<1/\Omega$ thus proving the analyticity
in $x$ of $\tilde u$ around $x=0$.

From the previous eq.(\ref{tildeequation}) we see that $\tilde
u(x,z_4)$ is analytic in the polydisk $|x|<1/\Omega$ and $z_4\in
D_4$. This assures not only that $u$ at infinity is bounded, a result
that we knew already from the treatment of sections
\ref{poincareprocedure}, \ref{linearsection} and Appendix C, but that
$\frac{\partial u}{\partial z_4}$ is uniformly bounded for all $z$
with $|z-z_4|>R_2$. Thus we have reproduced for $u$ the properties
P1-P5.

\bigskip

We have now to extend the properties P1-P5 to all the $u_k$ of
sections \ref{poincareprocedure} and \ref{linearsection} and to their
sum. First of all we notice that in solving equation
(\ref{lambdadiffeq}) the constants $c_k$ intervene.  These are given
by eq.(\ref{cequations}) i.e. by the ratio of two integrals where the
one which appears at the denominator never vanishes. The real analytic
dependence on $z_4$ of the denominator $\int\eta d^2$ is established
by the method provided in Appendix A while the derivative of the
numerator is again computed by splitting the integration region as
${\cal R}\cup{{\cal R}_c}$ and using the fact that $u_k$ and
$\frac{\partial u_k}{\partial z_4}, \frac{\partial\hat u_k}{\partial
  z_4}$ are bounded.

To establish the real analyticity of the sum of the series we shall
exploit the well known result that given a sequence of analytic
functions $f_n$ defined in a domain $\Omega$ which converge to $f$
uniformly on every compact subset of $\Omega$ then their sum is
analytic in $\Omega$ and the series of the derivatives $f'_n$ converge
uniformly to $f'$ on every compact subset of $\Omega$.

We saw in sections \ref{poincareprocedure} and \ref{linearsection}
that in general more than one extension step is required to reach the
complete solutions of eqs.(\ref{originaleq}) and
(\ref{generallinearequation}) but these steps are always finite in
number.  Let us consider first the case in which a single step is
sufficient

Then we have explicitly
\begin{eqnarray}
  && \Delta u= \theta e^u-\beta\label{basic2}\\
  &&u=\sum_{k=0}^\infty \lambda_1^k ~u_k\\
  && u_0=\log \alpha\\
  && \Delta u_1=\alpha\theta u_1 -\psi\\
  &&\Delta u_k=\alpha\theta(u_k+w_k),~~~~k\geq 2\label{nlinear2}\\
  && u_k=\sum_{h=0}^\infty \lambda_2^h~ u_{k,h}~.
\end{eqnarray}
Now we climb back the above sequence.  Starting from
\begin{eqnarray}\label{onesteplinear}
  &&\Delta u_{1,0}=-\varphi_0\nonumber\\
&&\Delta u_{1,1}= \eta(u_{1,0}+c_{1,0})-\psi\nonumber\\
  &&\Delta u_{1,2}= \eta (u_{1,1}+c_{1,1})\nonumber\\
  &&\Delta u_{1,3}= \eta (u_{1,2}+c_{1,2})\nonumber\\
  &&\dots
\end{eqnarray}  
and applying the inheritance result proven above we have that
$u_{1,h}$ are bounded in $D_4$ together with $\frac{\partial
  u_{1,h}}{\partial z_4}, \frac{\partial \hat u_{1,h}}{\partial z_4}$.
Being the convergence uniform we have that the sum
i.e. $u_1=\sum_{h=0}^\infty \lambda_2^h~ u_{1,h}$ is real analytic and
bounded in $D_4$ and that $\frac{\partial u_1}{\partial z_4},
\frac{\partial \hat u_1}{\partial z_4}$ are bounded in
$|z_4-z_4^0|<R_4-\frac{\varepsilon}{4}$.  The function $u_2$ is
obtained by solving (\ref{nlinear2}) where we recall that the source
$w_2$ depends only on $u_1$ and $w_k$ depends only on $u_r$
with $r<k$, in polynomial and thus analytic way.

Repeating the previous reasoning for $u_2$ we have analyticity and
boundedness of $u_2$ and its derivative in
$|z_4-z_4^0|<R_4-\frac{\varepsilon}{4}-\frac{\varepsilon}{8}$ and thus
boundedness of $u=\sum_{k=0}^\infty \lambda_1^k u_k$ in
$|z_4-z_4^0|<R_4-\frac{\varepsilon}{2}$ with its derivative bounded in
$|z_4-z_4^0|<R_4-\varepsilon$. Similarly one extends the analyticity
of $\hat u_k$ in $\zeta$ for $\zeta\neq0$ and of $u_k$ in $z$,
$|z-z_4|>R_2$ for $z\neq z_k$ to the sum of the series.
 
In the case the solution of eq.(\ref{basic2}) requires more
that one extension step one repeats the same procedure for each
extension step and the same for eq.(\ref{nlinear2}) keeping in mind
that such extension steps are always finite in number. Suppose
e.g. that the solution of eq.(\ref{basic2}) requires three
extension steps. Then we allocate for each step $\varepsilon/3$
instead of $\varepsilon$ and proceed as before. The same is done if
the intermediate linear equations require more than one extension
step. Here we employ the general result that given the equation
$\Delta u =\lambda \eta u - \phi_0-\lambda \psi$ if the sources
$\phi_0$ and $\psi$ are of the form $\eta s$ with $s$ having the
properties P1-P5 for $z_4\in D_4$, then the solution $u$ has the
properties P1-P5 for $|z_4-z_4^0|<R_4-\varepsilon$ for any
$\varepsilon>0$. Such a result is proven using exactly the treatment
of eq.(\ref{onesteplinear}) given above.

We recall now that the conformal field $\phi(z,z_4)$ is given in terms
of $u$ by
\begin{equation}
\phi(z,z_4)=u(z,z_4)+\nu(z,z_4)=u(z,z_4)-2\sum_k\eta_k\log|z-z_k|^2+ I(z,z_4)
\end{equation}  
where the analytic properties of $I(z,z_4)$ have already been given in
Appendix B. Thus we conclude that $\phi(z,z_4)$ is real analytic in
$z_4$ and in $z$ for $|z_4-z_4^0|< R_4-\varepsilon$ and for $z\neq z_k$.

\section{The real analyticity of the accessory parameters}
\label{realanalyticitysection}

Consider now a singularity $z_k$ with $k\neq 4$ and a circle $C$
around it of radius such that no other singularity is contained in
it. Given the conformal factor $\phi=u+\nu$ we have that both $u$ and
$\nu$ are real analytic functions of $z$ and $z_4$ for $z$ in an
annulus containing $C$ and $z_4\in D_4$. The accessory parameter $b_k$
can be expressed in terms of $\phi$ as
\begin{equation}\label{contourintegral}
b_k=\frac{1}{i\pi}\oint_{z_k} Q dz
\end{equation}  
where $Q$ is given by (see e.g. \cite{menottiHigherGenus})
\begin{equation}
  Q= -e^{\frac{\phi}{2}}\frac{\partial^2 }{\partial z^2}e^{-\frac{\phi}{2}}
  =\sum_k\frac{\eta_k(1-\eta_k)}{(z-z_k)^2}+
  \sum_k\frac{b_k}{2(z-z_k)}~.
\end{equation}  
Due to the analyticity of $\phi$ we can associate to any point of $C$ a
polydisk $D_z\times D_4$ where the $\phi$ is real analytic. Due to the
compactness of $C$ we can extract a finite covering provided by
such polydisks.

It follows then that the integral (\ref{contourintegral}) is a real
analytic function of $z_4$ for $z_4\in D_4$.  Thus we have that all
the accessory parameters $b_k$ with $k\neq 4$ are real analytic
functions of $z_4$. With regard to the accessory parameter $b_4$
we recall that due to the Fuchs relations \cite{menottiHigherGenus} it
is given in terms of the other $b_k$ and thus also $b_4$ is
real analytic in $z_4$. The reasoning obviously holds for the
dependence on the position of any singularity keeping the others
fixed, thus concluding the proof of the real analyticity on all source
positions.

We considered explicitly the case of the sphere topology with an
arbitrary number of elliptic singularities. This treatment extends the
results of \cite{menottiAccessory,menottiHigherGenus,menottiPreprint}
where it was found that in the case of the sphere with
four sources we had real analyticity almost everywhere. With the
almost everywhere attribute we could not exclude the occurrence of a
number of cusps in the dependence of the accessory parameters
on the position of the sources. Here we proved real analyticity
everywhere and for any number of sources and thus the occurrence of
cusps is excluded.  Obviously the whole reasoning holds when the
positions of the singularities are all distinct.  What happens when two
singularities meet has been studied only in special cases in
\cite{kra} and \cite{HS1,HS2}.

\section{Higher genus}\label{torusanalyticity}

For the case of the torus i.e. genus 1 we can follow the treatment of
\cite{menottiExistence}. 
In this case we know the explicit form of
the Green function
\begin{equation}
  G(z,z'|\tau)=\frac{1}{4\pi}\log[\theta_1(z-z'|\tau)\times c.c.]+ 
  \frac{i}{4(\tau-\bar\tau)}(z-z'-\bar z+\bar z')^2
\end{equation}  
where $\theta_1$ is the elliptic theta function \cite{batemanII}.
$G(z,z'|\tau)$ is a real analytic function in $z$, for $z\neq z'$,
and in $\tau$.
It satisfies
\begin{equation}
  \Delta  G(z,z'|\tau)=\delta^2(z-z') -  \frac{2i}{\tau-\bar\tau}~.
\end{equation}  

As for the $\beta$ we can construct it using the Weierstrass $\wp$
function.
\begin{equation}
\beta(z|\tau) = c\prod_k (\wp(z-z_k|\tau)\times c.c. )^{\eta_k}~.
\end{equation}
Using the freedom of $c$ we normalize the $\beta$ as to have
\begin{equation}
\int\beta(z|\tau)d^2z = 4\pi \sum_k 2\eta_k
\end{equation}
consistent with the topological restriction $\sum_k
2\eta_k>2(1-g)=0$. The $\nu$ is given by
\begin{equation}
  \nu = 4\pi\sum_k-2\eta_kG(z,z_k|\tau)+
  \int G(z,z'|\tau)\beta(z'|\tau)d^2z'
\end{equation}
with $\phi=u+\nu$.

We proceed now as in sections \ref{inheritance} and
\ref{realanalyticitysection} to obtain the real analytic dependence
of $\phi$ both on the position of the sources and on the modulus.
The $\phi(z)$ is translated to the
two sheet representation of the torus $\varphi(v,w)$ using
$v=\wp(z|\tau)$ and
\begin{equation}
\varphi(v,w) = \phi(z) + \log \big(\frac{d z}{d v}\times c.c.\big)
\end{equation}
where
\begin{equation}
  w = \frac{\partial v}{\partial z}
  =\wp'(z|\tau) =
  \sqrt{4(v-v_1)(v-v_2)(v-v_3)} 
\end{equation}
and thus
\begin{equation}
  \varphi(v,w) = \phi(z) -\frac{1}{2}
  \log[16(v-v_1)(\bar v-\bar v_1)
    (v-v_2)(\bar v-\bar v_2)(v-v_3)(\bar v-\bar v_3)]~.
\end{equation}
Now we proceed as in section \ref{realanalyticitysection}
where now \cite{menottiHigherGenus} in the auxiliary equation we have
\begin{eqnarray}\label{Qtorus}
  &&Q =\frac{3}{16}\bigg(\frac{1}{(v-v_1)^2}+\frac{1}{(v-v_2)^2}+
  \frac{1}{(v-v_3)^2}\bigg) \nonumber\\
&& +\frac{b_1}{2(v-v_1)}+\frac{b_2}{2(v-v_2)}+
\frac{b_3}{2(v-v_3)}\nonumber\\
&& + \sum_{k>3} \eta_k(1-\eta_k)\frac{(w+w_k)^2}{4(v-v_k)^2w^2}+
  \frac{b_k(w+w_k)}{4(v-v_k)w}~.
\end{eqnarray}
In eq.(\ref{Qtorus}) $w=\sqrt{4(v-v_1)(v-v_2)(v-v_3)}$ takes opposite
values on the two sheets and the factors $\frac{w+w_k}{2w}$ project the
singularities on the sheet to which they belong. We recall that the
accessory parameters $b_k$ are related by the Fuchs relations which in
the case of the torus are three in number \cite{menottiHigherGenus}
and thus the independent ones are as many as the sources. Then
proceeding as in the previous section we can extract by means of a
contour integral the real analytic dependence of the accessory parameters
on the source positions and on the modulus.

For higher genus we do not possess the explicit form of the Green
function and we have a representation of the analogue of the
Weierstrass $\wp$ function only for genus 2 \cite{komori}. Thus one
should employ more general arguments for the analyticity of the Green
function and for the expression of the $\beta$ function.

\section{Discussion and conclusions}\label{conclusions}

In the present paper we proved that on the sphere topology with any
number of elliptic singularities the accessory parameters are real
analytic functions of the source positions and the result has been
extended to the torus topology with any number of elliptic
singularities. This complements the result of Kra \cite{kra} where the
real analytic dependence was proven for parabolic and elliptic
singularities of finite order. Here the elliptic singularities are
completely general.  The extension of the present treatment to the
case when one or more singularities are parabolic should be in
principle feasible even though more complicated. Poincar\'e
\cite{poincare} in fact applied with success the continuation method
also in presence of parabolic singularities but the treatment is far
lengthier. The reason is that integrals of the type
\begin{equation}
\int \log|z-z'|^2 f(z')d^2z'
\end{equation}
with $f(z')$ behaving like
\begin{equation}
\frac{1}{|z'-z_k|^2 \log^2|z'-z_k|^2}  
\end{equation}
for $z'$ near $z_k$, diverge for $z\rightarrow z_k$, contrary to what
happens in the elliptic case. On the other hand it is proven in
\cite{poincare} that the solution of eq.(\ref{originaleq}) is finite
even at the parabolic singularities; in different words even if each
term of the series diverges for $z\rightarrow z_k$ their sum converges
to a function which is finite for $z\rightarrow z_k$, a procedure which
requires an higher number of iteration steps.  Some of these
intermediate steps employ $C^\infty$ but non analytic regularization.
This does not mean that the continuation method does not work for
parabolic singularities but simply that one one should revisit the
procedure of \cite{poincare} keeping analyticity in the forefront.

The real analytic dependence of the accessory parameter on the sphere
with four sources elliptic and/or parabolic and of the torus with one
source was proven already in
\cite{menottiAccessory,menottiHigherGenus,menottiPreprint} almost
everywhere in the moduli space using analytic variety
techniques. Almost everywhere meant e.g that we could not exclude the
presence of a number of cusps in the dependence of the
parameter on the source position or moduli. The results of the present
paper remove such a possibility.

In section \ref{torusanalyticity} we extended the procedure to the torus
topology in a rather straightforward way and thus in presence of $n$
elliptic sources on the torus we have that the independent accessory
parameters, which are $n$ in number, depend in real analytic way on
the source positions and on the modulus.

For higher genus i.e. $g>1$ the best approach appears to be the use of
the representation of the Riemann surface using the fuchsian domains
in the upper half-plane. For carrying through the program, in absence
of explicit forms of the Green function, one should establish its
analytic dependence on the moduli and also one should provide an
analytic $\beta$ satisfying the correct boundary conditions.

\bigskip

\section*{Appendix A}

In the text the problem arises to establish the
real analyticity of certain integrals. The problem can be dealt with
in two equivalent ways. The integral in question is a function of two
real variables $x,y$. To prove real analyticity we must show
that around a real point $x^0,y^0$ the function for real values of
$x,y$ is identical to the values taken by a holomorphic function
of two complex variables, call them $x^c,y^c$.

Alternatively one can use the complex variable $z$ and its complex
conjugate $\bar z$. Then proving the real analyticity of $f(z,\bar z)$
around $z^0,\bar z^0$ is equivalent to prove the analyticity of
$f(a,b)$ in $a$ and $b$ taken as independent variables. We shall use
this complex variable notation as it is simpler.

\bigskip

We prove here the real analyticity of $c(z_4)$ introduced in section
\ref{lichtenstein}. Write

\begin{equation}
  f(z,z_4)=\frac{\prod_{k\neq 4}[(z-z_k)(\bar z-\bar z_k)]^{-2\eta_k}}
  {[1+z\bar z]^{-2\sigma+2}}
[(z-z_4)(\bar z-\bar z_4)]^{-2\eta_4}\equiv g(z,\bar z) [(z-z_4)(z-z_4)]^{-2\eta_4}
\end{equation}
with $\sigma= \sum\eta_k$. In $g$ we ignored the dependence on
the $z_k$ with $k\neq 4$ as they will always be kept fixed.

In computing the derivative of $A$ defined by
\begin{equation}
    A=\int\frac{\prod_{k\neq 4}[(z-z_k)(\bar z-\bar
    z_k)]^{-2\eta_k}}{[1+z\bar z]^{-2\sigma+2}}
    [(z-z_4)(\bar z-\bar z_4)]^{-2\eta_4} \frac{i}{2}dz\wedge d\bar z
\end{equation}
w.r.t. $z_4$ in order to avoid the occurrence
of non integrable singularities it is expedient to apply the
technique of writing
\begin{equation}
A= A_1+A_2
\end{equation} 
being $A_1$ the integral extended inside a disk of center $z_4$ and
radius $R$ excluding all other singularities and $A_2$ the integral
outside.

Then we have
\begin{equation}
A_1=
\int_{\cal R} (\zeta\bar\zeta)^{-2\eta_4} g(\zeta+z_4,\bar\zeta+\bar z_4)
 \frac{i d\zeta\wedge d\bar \zeta}{2}
\end{equation}
which has derivative w.r.t. $z_4$
\begin{equation}\label{A1derivative}
\frac{\partial A_1}{\partial z_4}=
\int_{\cal R} (\zeta\bar\zeta)^{-2\eta_4}
\frac{\partial g(\zeta+z_4,\bar\zeta+\bar z_4)}{\partial z_4}
 \frac{i d\zeta\wedge d\bar\zeta}{2}~.
\end{equation}
It is justified to take the derivative operation inside the integral
sign as due to the real analyticity of $g$ in $\cal R$
the integrand in (\ref{A1derivative}) can be bounded by a function
$b(\zeta,\bar\zeta)$ independent of $z_4$ whose integral over $\cal R$
is absolutely convergent, exploiting $-2\eta_4+1>0$.

For $A_2$ we have
\begin{equation}
A_2 = \int_{{\cal R}_c} f(z,z_4) i \frac{d z\wedge d \bar z}{2}
\end{equation}
where ${\cal R}_c$ is the complement of $\cal R$, whose derivative is
given by
\begin{equation}\label{A2derivative}
\frac{\partial A_2}{\partial z_4}= \int_{{\cal R}_c} \frac{\partial
  f(z,z_4)}{\partial z_4 }\frac{i dz\wedge d\bar z}{2}
+\oint_{\partial {\cal R}_c} f(z,z_4) \frac{id\bar z}{2}~.
\end{equation}
Again it is legal to take the derivative operation inside the integral
sign in the first term of (\ref{A2derivative}) as we are working
outside $\cal R$ and the contour integral arises from the fact
that the domain ${\cal R}_c$ moves with $z_4$.

Such  term equals
\begin{equation}
i\oint f(z,z_4) \frac{dx}{2} + \oint f(z,z_4) \frac{dy}{2}~.
\end{equation}
We have also the complex conjugate equation which give rise to the
complex derivative w.r.t. $x,y$. Thus $c(z_4)=C(x_4,y_4)$ is
real analytic.

For future developments we point out the uniform bound for $|z|>2
~{\max} |z_k|$
\begin{equation}\label{generalbound}
  \beta=\frac{c(z_4)}{(1+z\bar z)^2}(1+\frac{1}{z\bar z})^{2\sigma}
  \prod_k\bigg|1-\frac{z_k}{z}\bigg|^{-4\eta_k}
  <\frac{{\rm const}}{(1+z\bar z)^2}(1+\frac{1}{z\bar z})^{2\sigma}
  \prod_k\bigg|1\pm\frac{1}{2}\bigg|^{-4\eta_k}
\end{equation}
with $+$ or $-$ according to $\eta_k<0$ or $\eta_k>0$ and for $z_4\in
D_4$.

\section*{Appendix B}
In this Appendix we work out the analytic properties of
\begin{equation}\label{Iintegral}
I(z,z_4)=\frac{1}{4\pi}\int\log|z-z'|^2 \beta(z',z_4) d^2z'
\end{equation}
which are necessary to establish the properties of the function
$\theta(z,z_4)$.
We have
\begin{equation}
  I(z,z_4)=\frac{1}{4\pi}\log (z\bar z)\int\beta(z',z_4) d^2z'+
  \frac{1}{4\pi}\int\log|1-\frac{z'}{z}|^2 \beta(z',z_4) d^2z'.
\end{equation}
Let $\Omega$ be the radius of a disk which encloses all singularities
$z_k$; outside such a disk we have
\begin{equation}
 \beta(z,z_4)<\frac{c}{(z\bar z)^2}
\end{equation}
with $c$ independent of $z_4$ for $z_4\in D_4$.
Moreover we choose $\Omega>1$. We have 
\begin{eqnarray}
&&  \int\log|1-\frac{z'}{z}|^2 \beta(z',z_4) d^2z'=\\
&&  |z|^2\int_{|y|<\frac{1}{2}}\log|1-y|^2 \beta(zy,z_4) d^2y+
  |z|^2\int_{|y|>\frac{1}{2}}\log|1-y|^2 \beta(zy,z_4) d^2y
\end{eqnarray}
First we examine the region $|z|>2\Omega$. The first integral
is less than
\begin{equation}
  2\log 2~|z|^2\int_{|y|<\frac{1}{2}}\beta(zy,z_4)d^2y\leq
  2\log 2\int\beta(z',z_4)d^2z'
\end{equation}  
and the second, due to $\beta(z,z_4)<c/(z\bar z)^2$ is less than
\begin{equation}
  \frac{c}{z\bar z}\int_{|y|>\frac{1}{2}}
  \big|\log|1-y|^2\big|\frac{d^2y}{(y\bar y)^2}~.
\end{equation}  
Thus for $|z|>2\Omega$ we have
\begin{eqnarray}
  &&  |I(z,z_4)|\leq\frac{1}{4\pi}\log(z\bar z)\int\beta(z',z_4) d^2z'
  + \frac{2 \log 2}{4\pi} ~\int\beta(z',z_4) d^2z'\\
&&+ \frac{c}{4\pi z\bar z}\int_{|y|>\frac{1}{2}}
  \big|\log|1-y|^2\big|\frac{d^2y}{(y\bar y)^2}~.
\end{eqnarray}
For $|z|<2\Omega$ we isolate the singularities $z_k$ of $\beta$ by non
overlapping discs of radius $a<\frac{1}{4}$. In the complement
$\beta(z,z_4)$ is majorized by $c/(1+z\bar z)^2$ with $c$ independent
of $z_4$ for $z_4\in D_4$. We bound $|I|$ by the sum of two
terms the first being 
\begin{eqnarray}
&&  \frac{c}{4\pi}\int_{|\zeta|>1} \log\zeta\bar\zeta\frac{1}{(1+(z+\zeta)
    (\bar z+\bar\zeta))^2}d^2\zeta\leq\\
&&  \frac{c}{4\pi}\int_{1<|\zeta|<2\Omega}\log\zeta\bar\zeta ~d^2\zeta+
  \frac{c}{4\pi}\int_{|\zeta|>2\Omega}\log\zeta\bar\zeta
  \frac{1}{(1+(|\zeta|-2\Omega)^2)^2}d^2\zeta
\end{eqnarray}  
and the second is the contribution of $|\zeta|<1$, where
$\log\zeta\bar\zeta$ is negative
\begin{eqnarray}
-\frac{c}{4\pi}\int_{|\zeta|<1} \log\zeta\bar\zeta\frac{1}{(1+(z+\zeta)
  (\bar z+\bar\zeta))^2}d^2\zeta\leq
-\frac{c}{4\pi}\int_{|\zeta|<1}\log\zeta\bar\zeta ~d^2\zeta~.
\end{eqnarray}  

The singularity at $z'=z_k$ is dealt with $\zeta=z-z_k$. The contribution
of the disk of radius $a$ is
\begin{equation}
  4\pi I_a=  \int_a\log|\zeta-\zeta'|^2\tilde\beta(\zeta',z_4)d^2\zeta'
\end{equation}  
where $\tilde\beta(\zeta,z_4)=\beta(\zeta+z_k,z_4)$.

We have
\begin{equation}
4\pi I_a= \log|\zeta|^2 \int_{a} \tilde\beta(\zeta',z_4)d\zeta'+ \int_{a}
  \log\bigg|1-\frac{\zeta'}{\zeta}\bigg|^2\tilde\beta(\zeta,z_4) \hat
d^2\zeta'
\end{equation}
and thus for $|\zeta|>2 a$ we have
\begin{equation}
  |4\pi I_a|\leq (\big|\log|\zeta|^2\big|+2\log 2) \int_{a}
  \tilde\beta(\zeta',z_4)d\zeta'
\end{equation}
and as we are working for $|z|<2\Omega$ the $|\log|\zeta|^2|$
is bounded.
For $a<|\zeta|<2a$ as $\log|\zeta-\zeta'|^2$ is always
negative we have
\begin{equation}
  |4\pi I_a|\leq -M\int_{a} \log|\zeta-\zeta'|^2
  (\zeta'\bar\zeta')^{-2\eta_k} d^2\zeta' =
-\pi M\log(\zeta\bar\zeta)\frac{(a^2)^{1-2\eta_k}}{1-2\eta_k}
\end{equation}
where $M$ is such that
$\tilde\beta(\zeta',z_4)<M(\zeta'\bar\zeta')^{-2\eta_k}$ for $\zeta$
in the disk of radius $a$ and $z_4\in D_4$.

Finally for $|\zeta|<a$ 
\begin{equation}
 |4\pi I_a|\leq
  \frac{\pi M}{1-2\eta_k}\bigg[-(a^2)^{1-2\eta_k}\log a^2
    +\frac{(a^2)^{1-2\eta_k}-(\zeta\bar\zeta)^{1-2\eta_k}}
    {1-2\eta_k}\bigg]~.
\end{equation}

We conclude that $I(z,z_4)$ for any $z$ and $z_4\in D_4$ is always finite
and bounded by
\begin{equation}\label{summaryI}
 |I(z,z_4)|\leq\frac{1}{4\pi}\log (z\bar z+1)\int\beta(z,z_4)d^2z+ c_1
\end{equation}
with $c_1$ independent of $z_4$ for $z_4\in D_4$.

We prove now that $I(z,z_4)$ is analytic in $z$ for $z\neq z_k$.
Given a $z_0\neq z_k$ let us consider a disk $D$ of center $z_0$ and
radius $r$ such that such disk does not contain any singularity of
$\beta(z,z_4)$. By standard arguments one shows that the derivative
w.r.t. $z$ of the contribution of such disk to the integral
(\ref{Iintegral}) is
\begin{equation}\label{Dintegral}
\frac{1}{4\pi} \int_D\frac{\beta(z',z_4)}{z-z'}d^2z~.
\end{equation}
The contribution of the complement $D_c$ of $D$ to the derivative is
\begin{equation}\label{Dcintegral}
\frac{1}{4\pi} \int_{D_c}\frac{\beta(z',z_4)}{z-z'}d^2z
\end{equation}
as for $|z-z_0|<\frac{r}{2}$ we have that the integrand
is bounded by
\begin{equation}
\frac{\beta(z',z_4)}{|z'-z_0|-\frac{r}{2}}
\end{equation}
which is absolutely convergent and independent of $z$.
Thus $I(z,z_4)$ is analytic in $z$ for $z\neq z_k$ and its derivative
is given by the sum of (\ref{Dintegral}) and (\ref{Dcintegral}) i.e.
by  (\ref{Dintegral}) with $D$ replaced by the whole $z$ plane.

In working out the derivative of $I(z,z_4)$ w.r.t. $z_4$ in order to
avoid non integrable singularities we must isolate a disk ${\cal R}_1$
of fixed radius $R_1$ with center $z_4$ and excluding all other $z_k$.

Thus as given in section \ref{inheritance} we write for $|\zeta|<R_1$,
$\zeta=z-z_4$,
\begin{eqnarray}\label{IR1}
\hat I(\zeta,z_4)&=&\frac{1}{4\pi}\int_{{\cal R}_1}
  \log|\zeta-\zeta'|^2 \hat\beta(\zeta',z_4)d^2\zeta'\nonumber\\
&+&\frac{1}{4\pi}\int_{{\cal R}_{1c}}
    \log|\zeta+z_4-z'|^2  \beta(z',z_4)d^2z'~.
\end{eqnarray}
where ${\cal R}_{1c}$ is the complement of ${\cal R}_1$.

We notice that $\hat I(\zeta,z_4)$ does not depend on the specific
choice of the radius $R_1$ of the domain used in (\ref{IR1}) and in
(\ref{derivativeofhatI}) below to compute the derivative w.r.t. $z_4$,
provided that ${\cal R}_1$ does not contain any other singularity
except $z_4$.

Its derivative w.r.t. $z_4$ is
\begin{eqnarray}\label{derivativeofhatI} 
\frac{\partial \hat I(\zeta,z_4)}{\partial z_4}
  &=&\frac{1}{4\pi}\int_{{\cal R}_1}
  \log|\zeta-\zeta'|^2 \frac{\partial \hat\beta(\zeta',z_4)}
                    {\partial x_4 }d^2\zeta\nonumber\\
      &+&\frac{1}{4\pi}\int_{{\cal R}_{1c}}
  \frac{1}{\zeta+z_4-z'}\beta(z',z_4) d^2 z'\nonumber\\
 &+&\frac{1}{4\pi}\int_{{\cal R}_{1c}} \log|\zeta+z_4-z'|^2
  \frac{\partial \beta(z',z_4)}{\partial z_4}d^2 z'\nonumber\\
  &+&\frac{i}{8\pi}\oint_{\partial {\cal R}_{1c}}
  \log|\zeta+z_4-z'|^2 \beta(z',z_4) d\bar z'~.
\end{eqnarray}
The contour integral is the contribution of the dependence of the
domain ${\cal R}_{1c}$ on $z_4$.  In the first term of
(\ref{derivativeofhatI}) taking the derivative under the integral sign
is legal because the integrand is of the form
$(\zeta\bar\zeta)^{-2\eta_4} \frac{\partial f(\zeta,z_4)}{\partial
  z_4}$ and this expression can be majorized for $|\zeta|<R_1$ by a
function independent of $z_4$ for $z_4\in D_4$ whose integral is
absolutely convergent due to $-2\eta_4+1>0$.  In the second
term the denominator $\zeta+z_4-z'$ never vanishes and we can apply
the same majorization. In the third term the $\log$ is not singular;
$\frac{\partial \beta(z',z_4)}{\partial z_4}$ has the singularity
$(|z'-z_4|^2)^{-2\eta_4}/(z'-z_4)$ which is non integrable for
$-4\eta_4 +1< 0$ but such a singularity lies outside the integration
region ${\cal R}_{1c}$. As for the contour integral, on it both the
logarithm and the $\beta$ are regular.

We consider then an other disk ${\cal R}_2$ centered again in $z_4$ and
of radius $R_2<R_1$.
For $|z-z_4|>R_2$ we use the expression
\begin{eqnarray}
I(z,z_4)&=&\frac{1}{4\pi}\int_{{\cal R}_2}
  \log|z-\zeta'-z_4|^2 \hat\beta(\zeta',z_4)d^2\zeta'\nonumber\\
&+& \frac{1}{4\pi}\int_{{\cal R}_{2c}}
    \log|z-z'|^2  \beta(z',z_4)d^2z'~.
\end{eqnarray}
Its derivative w.r.t. $z_4$ is given by
\begin{eqnarray}\label{derivativeofI}
&&-\frac{1}{4\pi}\int_{{\cal R}_2}
\frac{1}{z-\zeta'-z_4} \hat\beta(\zeta',z_4)d^2\zeta'\nonumber\\
&& +\frac{1}{4\pi}\int_{{\cal R}_2}
  \log|z-\zeta'-z_4|^2 \frac{\partial
    \hat\beta(\zeta',z_4)}{\partial z_4} d^2\zeta'\nonumber\\
&& +\frac{1}{4\pi}\oint_{{\cal R}_{2c}}
    \log|z-z'|^2 \frac{\partial
    \beta(z',z_4)}{\partial z_4}d\bar z'\nonumber\\
&& +\frac{i}{8\pi}\oint_{\partial {{\cal R}_2c}}
    \log|z-z'|^2 \beta(z',z_4) d\bar z'~.
\end{eqnarray}

Regarding the first two terms in (\ref{derivativeofI}) as $z-\zeta'-z_4$
never vanishes in ${\cal R}_2$ it is legal to take the derivative under
the integral sign. In the third term the integrand can be majorized by
a $z_4$ independent integrable function. The last term is the
contribution of the moving integration region.
Thus we have analyticity of $I(z,z_4)$ for $|z-z_4|>R_2$, $z_4\in D_4$.

We conclude that $I(z,z_4)$ is everywhere finite, bounded by
(\ref{summaryI}), continuous in $z, z_4$.  $I(z,z_4)$ is analytic for
$z\neq z_k$.  $\hat I(\zeta,z_4)$ is analytic in $z_4$ for $z_4 \in
D_4$ $|\zeta|<R_1$ while $I(z,z_4)$ is analytic in $z_4$ for $z_4\in
D_4$, $|z-z_4|>R_2$.

In the text we shall also need some information about the behavior of
$I(z,z_4)$ and $\frac{\partial I(z,z_4)}{\partial z_4}$ for large $z$. We
already saw that $I(z,z_4)$ behaves at infinity like $(\sum_k 2\eta_k
-2)\log|z|^2 $ and from (\ref{derivativeofI}) we have the simple bound
$|\frac{\partial I(z,z_4)}{\partial z_4}|< {\rm const}~\log|z|^2$.

\section*{Appendix C}

The main tool used in the text is the solution of the
equation
\begin{equation}\label{sourceequation}
  \Delta u(z,z_4) = \theta(z,z_4) s(z,z_4)
\end{equation}
under the condition on the source
\begin{equation}\label{zerointcondition}
    \int \theta(z,z_4)s(z,z_4)) d^2z=0~.
\end{equation}
The solution of eq.(\ref{sourceequation}) apart for the addition of an harmonic
function is
\begin{equation}\label{sourcesolution}
\frac{1}{4\pi}\int\log|z-z'|^2 \theta(z',z_4) s(z',z_4) d^2z'
\end{equation}
and as we are interested in bounded solutions the only freedom will be the
addition of a constant. The purpose here is to give a bound on
(\ref{sourcesolution}).

We recall that
\begin{equation}
\theta(z,z_4)=\prod_k [(z-z_k)(\bar z -\bar z_k)]^{-2\eta_k} ~e^{I(z,z_4)}=
\beta(z,z_4) ~r(z,z_4)
\end{equation}
with $0<r_1<r<r_2$.

The function $\theta(z,z_4)$ is positive with elliptic singularities
and bounded at infinity by $\frac{\rm const}{|z\bar z|^2}$ and we
shall give a bound on (\ref{sourcesolution}) in terms of
$\max|s(z,z_4)|$. As $r_1<r<r_2$ most of the techniques for proving
such a bound have been already worked out in the preceding Appendix B.

We have due to the condition
(\ref{zerointcondition}) and the bound (\ref{generalbound})
\begin{equation}\label{thetwoforms}
  \int\log|z-z'|^2 \theta(z',z_4)s(z',z_4)d^2z'=
  \int\log|1-\frac{z'}{z}|^2 \theta(z',z_4)s(z',z_4)d^2z'~.
\end{equation}
For $|z|>2\Omega$ we use the second form in eq.(\ref{thetwoforms})
replacing $s(z',z_4)$ with $\max |s(z',z_4)|$ and the logarithm by its
absolute value and the proceeding as in the previous Appendix. It is
important to notice that now due to the condition
(\ref{zerointcondition}) the term $\log z\bar z$ which diverges at
infinity in eq.(\ref{summaryI}) is absent and thus we have boundedness
also at infinity. The region $|z|<2\Omega$ is treated exactly as in the
previous Appendix B. The result is that
\begin{equation}
  \bigg|\frac{1}{4\pi} \int\log|z-z'|^2 \theta(z',z_4)s(z',z_4)d^2z'\bigg|
  < B ~\max|s(z,z_4)|
\end{equation}
with $B$ independent of $z_4$ for $z_4\in D_4$.

\bigskip

\section*{Appendix D}

In this appendix we report the proof that the series (\ref{nonlinearseries})
converges with a non zero convergence radius.

From the text we have for $k\geq 2$
\begin{equation}
\max|u_k|\leq\max|w_k|~.
\end{equation}  
Given  $\max|u_1| \equiv \gamma_1$ one considers the series
\begin{equation}
\nu = \lambda \gamma_1+\lambda^2 \gamma_2+\lambda^3 \gamma_3+\dots
\end{equation}  
where (see section \ref{poincareprocedure})
\begin{equation}
  \gamma_2= \frac{\gamma_1^2}{2},~
  \gamma_3= \frac{\gamma_1^3}{6}+\gamma_1\gamma_2,~
  \gamma_4 = \frac{\gamma_1^4}{24}+\frac{\gamma_1^2 \gamma_2+\gamma_2^2}{2}
  +\gamma_1\gamma_3,~~~~\dots
  \dots
\end{equation}  
Obviously such a series of positive terms majorizes term by term
the series $\lambda u_1+\lambda^2 u_2+\lambda^3 u_3+\dots$. We want to find
the convergence radius of $\nu$. 
For the function $\nu$ we have
\begin{eqnarray}
e^\nu = 1+\lambda \gamma_1&+&\lambda^2 \gamma_2+\lambda^3 \gamma_3
  +\dots\nonumber \\
  &+&\lambda^2 \gamma_2+\lambda^3 \gamma_3+\dots \nonumber\\
 &=&1+2\nu-\lambda \gamma_1~.
\end{eqnarray}
Let us consider the implicit function defined by
\begin{equation}
1+2 \nu - e^\nu = \lambda \gamma_1~.
\end{equation}  
As for $\nu=0$ we have $\lambda \gamma_1=0$ and 
\begin{equation}\label{jacobian}
\frac{\partial(1+2\nu-e^\nu)}{\partial \nu}= 2-e^\nu
\end{equation}
which equals $1$ at $\nu=0$ the analytic implicit function theorem
assures us the the $\nu$ is analytic in $\lambda$ in a finite disk
around $\lambda=0$ and thus with a non zero radius of convergence in
the power expansion.  This suffices for the developments of the
present paper. Such radius of convergence $r_0$ can be computed
\cite{poincare} and is given by $r_0=(\log4-1)/\gamma_1$, corresponding
to the vanishing of (\ref{jacobian}).


\bigskip


\begin{thebibliography}{99}

\bibitem{bolibrukh} A.A. Bolibrukh, {\it The Riemann-Hilbert problem}, Russian
  Math. Surveys 45:2 (1990) 1-47, Uspekhi Mat. Nauk 45:2 (1990) 3-47

\bibitem{CMS2} L. Cantini, P. Menotti, D. Seminara, {\it Liouville
  theory, accessory parameters and (2+1)-dimensional gravity},
  Nucl. Phys. B638 (2002) 351

\bibitem{ZT1} P.G. Zograf and L.A. Takhtajan, {\it On Liouville equation,
  accessory parameters, and the geometry of Teichm\"uller space for
  Riemann surfaces of genus $0$}, Math. USSR Sbornik 60 (1988) 143

\bibitem{ZT2} P.G. Zograf and L.A. Takhtajan, {\it On uniformization
  of Riemann surfaces and the Weyl-Peterson metric on Teichm\"uller
  and Schottky spaces}, Math. USSR Sbornik vol.60 (1988) 297

\bibitem{CMS1} L. Cantini, P. Menotti, D. Seminara, {\it  Proof of
  Polyakov conjecture for general elliptic singularities}, 
  Phys. Lett. B517 (2001) 203

\bibitem{TZ} L. A. Takhtajan, P. G. Zograf, {\it Hyperbolic 2 spheres
  with conical singularities, accessory parameters and K\"ahler metrics
  on M(0,n)}, Trans. Am.  Math. Soc. 355 (2003) 1857

\bibitem{ZZ} A. Zamolodchikov and Al. Zamolodchikov, {\it Structure
  constants and conformal bootstrap in Liouville field theory},
  Nucl. Phys. B 477 (1996) 577

\bibitem{HJP} L. Hadasz, Z.Jaskolsy and M. Piatek, {\it Classical geometry
from quantum Liouville theory}, Nucl. Phys. B 724 (2005) 529

  
\bibitem{FP} F. Ferrari and M. Piatek, {\it Liouville theory, N=2
gauge theories and accessory parameters}, JHEP 05 (2012) 025

\bibitem{menottiAccessory} P. Menotti, {\it  Accessory parameters for
  Liouville theory on the torus}, JHEP 12 (2012) 001

\bibitem{piatek} M. Piatek, {\it Classical torus conformal blocks,
  $N=2^*$ twisted superpotential and the accessory parameter of
Lam\'e equation}, JHEP 03 (2014) 124

\bibitem{LLNZ} A. Litvinov, S. Lukyanov, N. Nekrasov,
  A. Zamolodchikov, {\it Classical conformal blocks and Painlev\'e'
    VI}, JHEP 07 (2014) 144

\bibitem{menottiConformalBlocks} P. Menotti, {\it Classical
  conformal blocks}, Mod. Phys. Lett. A 31 (2016) 27, 1650159

\bibitem{menottiTorusBlocks} P. Menotti, {\it Torus classical
  conformal blocks}, Mod. Phys. Lett. A33 (2018) 28, 1805.07788

\bibitem{kra} I. Kra, {\it Accessory parameters for punctured spheres}, 
Trans. Am. Math. Soc. 313 (1989) 589


\bibitem{menottiPreprint} P. Menotti, {\it Real analyticity of
  accessory parameters}, e-Print: 2002.09933 [hep-th]

  
\bibitem{menottiHigherGenus} P. Menotti, {\it The Polyakov relation for the
  sphere and higher genus surfaces}, J. Phys. A49 (2016)19, 195203
  
\bibitem{poincare} H. Poincar\'e, {\it Les functions fuchsiennes et
  l'equation $\Delta u=e^u$}, J. Math. Pures Appl. t.4 (1898) 137

\bibitem{lichtenstein} L. Lichtenstein, {\it Integration der
  Differentialgleichung $\Delta_2u=ke^u$ auf geschlossenen Fl\"achen},
  Acta mathematica 40 (1915) 1

\bibitem{troyanov} M. Troyanov, {\it Prescribing curvature on compact
  surfaces with conical singularities}, Trans. Am. Math. Soc. 324
  (1991) 793

\bibitem{menottiExistence} P. Menotti, {\it On the solution of the
  Liouville equation}, J. Phys. A50 (2017) 37, 375205

\bibitem{leroy} E. Le Roy, {\it Sur la determination des integrales de
  certain equation aux derrivee partielle non lineaires par leur
  valeurs sur un surface fermee}, C.R.A.S. Paris 124 (1897) 1508

\bibitem{mahwin} J. Mahwin, {\it Henry Poincar\'e and partial
  differential equations},  Nieuw. Arch. Wiskd. (2012) 159

\bibitem{HS1} J.A. Hempel and S.J. Smith, {\it Uniformization of the
  twice punctured disk- Problems of confluence},
  Bull. Austral. Math. Soc. vol.39 (1989) 369

\bibitem{HS2} J. A. Hempel and S.J. Smith, {\it Hyperbolic lengths
  of geodesics surrounding two punctures},
  Proc. Am. Math. Soc.  vol.103 (1988) 513

\bibitem{batemanII} A. Erdelyi, {\it Higher Transcendental Functions}
  vol.2, McGraw-Hill (New York) 1953

\bibitem{komori} Y. Komori, {\it On the automorphic functions for
  Fuchsian groups of genus two}, Spaces of Kleinian Groups, London
  Math. Soc. Lect. Notes vol.329, Cambridge University
  Press (2006) 259

 
\end{thebibliography}
\end{document}